\documentclass[journal]{IEEEtran}

\usepackage{cite}

\ifCLASSINFOpdf
  \usepackage[pdftex]{graphicx}
\else
  \usepackage[dvips]{graphicx}
\fi

\usepackage[cmex10]{amsmath}
\usepackage{amssymb}
\usepackage{float}

\usepackage{bm}
\usepackage{extarrows}

\usepackage{algorithm}
\usepackage{algorithmic}

\usepackage{xcolor}
\usepackage{subfigure}

\usepackage{stfloats}

\makeatletter

\newcommand{\Rmnum}[1]{\expandafter\@slowromancap\romannumeral #1@}
\makeatother

\usepackage{footnote}
\makesavenoteenv{tabular}

\begin{document}
\vspace{-0.5cm}
\title{ Study of Unique-Word Based GFDM Transmission Systems}

\author{\IEEEauthorblockN{Joao T. Dias$^{1,2}$ and 
Rodrigo C. de Lamare$^{2,3}$} \\
\vspace{5mm}
\IEEEauthorblockA{$^{1}$Federal Center for Technological Education of Rio de Janeiro (CEFET-RJ)},\\
$^{2}$Centre for Telecommunications Studies (CETUC) of Pontifical Catholic University of Rio de Janeiro, Brazil, and\\
$^{3}$Department of Electronic Engineering, University of York, UK\\
e-mail: joao.dias@cefet-rj.br, $\left\{ \text{joao.dias, delamare} \right\}$@cetuc.puc-rio.br and rodrigo.delamare@york.ac.uk}
\maketitle

\begin{abstract} ~

In this paper, we propose  the use of a deterministic sequence,
known as unique word (UW), instead of the cyclic prefix (CP) in
generalized frequency division multiplexing (GFDM) systems. The UW
consists of known sequences that,  if not null, can be used
advantageously for synchronization and channel estimation purposes.
In addition, UW allows the application of a highly efficient linear
minimum mean squared error (LMMSE) smoother for noise reduction at
the receiver.  To avoid the conditions of non-orthogonality caused
by the insertion of the UW and performance degradation in time
varying frequency-selective channels, we use frequency-shift offset
quadrature amplitude modulation (FS-OQAM). We present a signal model
of a UW-GFDM system considering a single and multiple UWs. We then
develop an LMMSE receive filter for signal reception of the proposed
UW-GFDM system. Simulations show that the proposed UW-GFDM system
outperforms  prior work.
\end{abstract}


\IEEEpeerreviewmaketitle

\section{Introduction}

In recent years, there has been an increase in the demand for mobile
communication systems and the evolution of these systems have
focused, as a priority, on the increase in throughput. However, in
scenarios predicted for the future generation of mobile
communications, such as machine-to-machine communication (M2M),
Internet of Things (IoT), tactile Internet and wireless regional
area networks (WRAN), there are characteristics that clearly go
beyond the high data rates~\cite{Nicola}. The challenges posed by
these scenarios require: low power consumption, which makes it a
problem for the synchronization of orthogonal frequency division
multiplexing (OFDM) systems to maintain orthogonality between the
sub-carriers; short bursts of data, which prohibit the use of a
cyclic prefix (CP) in all symbols due to low spectral efficiency;
and the high out-of-band (OOB) emission of OFDM that is an issue for
dynamic and opportunistic spectrum access.

Due to these requirements, Generalized Frequency Division
Multiplexing (GFDM) \cite{Nicola} has been proposed for the air
interface of 5G networks. The flexibility of GFDM allows it to cover
CP-OFDM as a special case. However, filtering of subcarriers results
in non-orthogonal waveforms, inter-symbol interference (ISI) and
inter-carrier interference (ICI). The Unique Word (UW) concept has
been proposed for OFDM in~\cite{Huemer} along with an optimized
receiver concept. Since CP is random, its only function is to avoid
the interference caused by the channel delay. The UW, being
deterministic, if a non null UW is used, can be used also for
synchronization and channel estimation. Many studies with UW have
shown that a significant gain in bit error rate (BER) over OFDM can
be obtained~\cite{Huemer, Onic, Huber}.

In this work, we propose the use of UW in GFDM systems which results
in UW-GFDM systems. We devise a signal model of a UW-GFDM system
considering a single UW and multiple UWs. We then develop an LMMSE
receive filter for signal reception of the proposed UW-GFDM system.
Simulations show that the proposed UW-GFDM system outperforms prior
work. The paper is organized as follows: in Section II, a
description of the signal model of the UW-GFDM system is given;
Section III describes the UW-GFDM system design; the results of the
simulations are presented in Section IV and in the Section V, some
conclusions are drawn.

\section{Proposed Signal Model}

In the proposed UW-GFDM systems, the input to the system is a binary
sequence organized in a vector $\bold{b}$ of length $\mu N_d M$,
where $\mu$ is the order of modulation, $N_d$ is the length of the
data vector used in the GFDM sub-symbol and $M$ is the number of
time slots that make up one GFDM block.

The first step is the mapper, where the binary sequence is mapped
into a complex valued sequence $\bold{d}_d$, according to the
mapping scheme adopted. In this work, we use the frequency-shift
offset quadrature amplitude modulation (FS-OQAM)~\cite{Gaspar} that
can be exploited to address non-orthogonal conditions. Afterwards,
the sequence of length ${N}_d$ obtained with the modulation of each
group of $\mu N_d$ bits is added to $N_r$ redundant subcarriers,  so
that $N_d + N_r = K$, where $K$ is the number of subcarriers per
GFDM sub-symbol.

Then, the resulting sequence $\bold{d}$ of length $MK$ is reshaped
by serial-to-parallel conversion to a matrix $\bold{D}$ with
dimensions $M\times K$. To obtain interference-free transmission,
each element $d_{k,m}$ of the matrix $\bold{D}$ is transmitted with
its real and imaginary part $d_{k,m} = d_{k,m}^{(i)} +
jd_{k,m}^{(q)}$ using symmetric, real-valued, half Nyquist prototype
filter with the offset of $K/2$ samples from each other and phase
rotation of $\frac{\pi}{2}$ radians among adjacent subcarriers and
sub-symbols, given by
\begin{equation} \label{eq:filter}
\begin{array}{lr}
g_{k,m}^{(i)} [n] = j^{k} g[(n-mK)~~ \text{mod}~N] e^{j2\pi \frac{k}{K}n},\\
g_{k,m}^{(q)} [n] = j^{k+1} g[(n-(m+\frac{1}{2})K)~~ \text{mod}~N] e^{j2\pi \frac{k}{K}n}.
\end{array}
\end{equation}
where $n=0, 1,..., N-1$ is the time index and $N=KM$.


The resulting GFDM signal can be described by 
\begin{equation} \label{eq:gfdmtransmit}
x [n]=\displaystyle \sum_{m=0}^{M-1}\displaystyle \sum_{k=0}^{K-1} d_{k,m}^{(i)} g_{k,m}^{(i)} [n] + \displaystyle \sum_{m=0}^{M-1}\displaystyle \sum_{k=0}^{K-1} d_{k,m}^{(q)} g_{k,m}^{(q)} [n].
\end{equation}

Before transmitting the signal a UW with length $L$ or multiple UWs
are added in order to preserve the circulant structure  and to make
frequency domain equalization possible at the receiver. Afterwards,
the signal is converted to the analog domain and transmitted.

At the receiver, the received signal is described by
\begin{equation} \label{eq:signalreceived}
y' (n)= x'(n)\ast h(n)  + \omega (n) 
\end{equation}
where $h(n)$, denotes the impulse response of the channel, $x'(n)$ is the transmitted signal with addition of the UW, and $\omega (n)$ is a additive white Gaussian noise.

At the receiver, the signal is converted from the analog to the digital domain, followed by the removal of the UW and equalized by
\begin{equation} 
y(n)= \text{IFFT} \left\{\frac{\text{FFT}[y'(n)]}{\text{FFT}[h(n)]}\right\} 
\end{equation}

The received data can be described by
\begin{equation} 
\widehat{d}_{k,m} = \widehat{d}_{k,m}^{(i)} + j\widehat{d}_{k,m}^{(q)}
\end{equation}
and,
\begin{equation} 
 \widehat{d}_{k,m}^{(.)}= \Re \left\{y(n)\circledast g_{k,m}^{(.)\ast} [-n]\right\} \mid_{n=0},\\
\end{equation}
where $\circledast$ denotes circular convolution with period $N$, and $^{(.)}$ represents $^{(i)}$ or $^{(q)}$.

\begin{figure*}
\centering
\includegraphics[width=.9\textwidth]{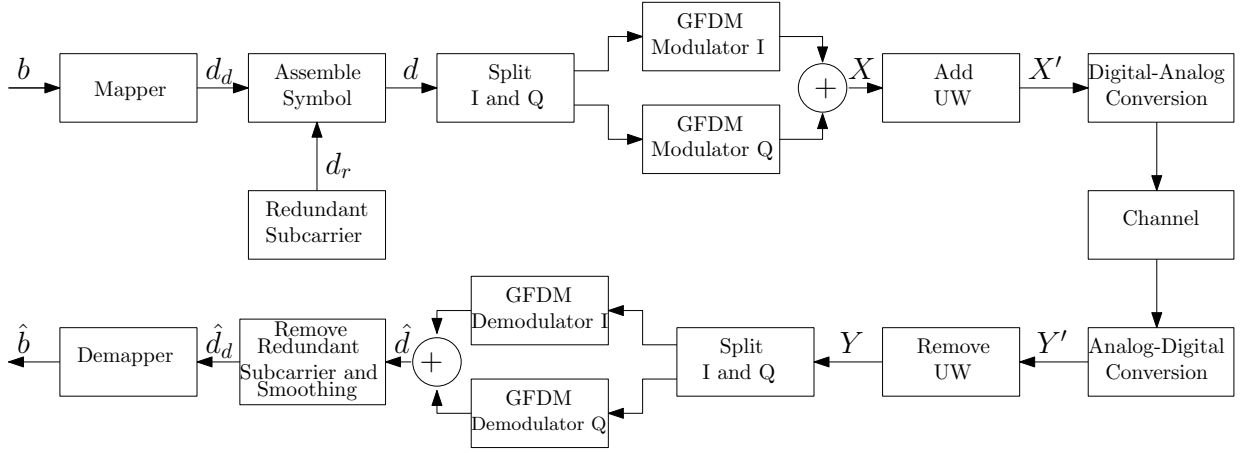}
\caption{GFDM block diagram}
\label{fig:GFDMblockdiagram}
\end{figure*}

The received symbol vector $\widehat{\bold{d}}_d$ is obtained by
applying the parallel to serial converter, the Wiener smoothing
operation and extracting the data part of the signal
$\widehat{\bold{d}}$. After that, the data part of the signal
$\widehat{\bold{d}}_d$ is demapped into the detected bit vector
$\widehat{\bold{b}}$. The block diagram of the proposed UW-GFDM
system is shown in fig.~\ref{fig:GFDMblockdiagram}.

\section{Design of the UW-GFDM System}

In this section, we detail the design of UW-GFDM systems \cite{uwgfdm}. The framing of UW is made by introducing a predefined sequence $\bold{x}_u$ which shall form the tail of the data vector in the time domain ${\bold{x'}} = [\bold{x}_d ^T~~\bold{x}_u ^T]$. 
Therefore, the sequence $\bold{x}$ assumes the form $\bold{x} =
[\bold{x}_d ^T~~\bold{0}^T]$ aiming to add the UW of length L in the
time domain and to obtain $\bold{x'} = \bold{x} +
[\bold{0}^T~~\bold{x}_u ^T]$ \cite{Onic}.

In the first step, in order to get the vector structure $\bold{x}$
in time domain, a reduction of the number $N_d$ of data subcarriers
is made and a set of redundant subcarriers $N_r$ is added, in the
frequency domain.
In case of multiple UWs in GFDM systems, the matrix $\bold{D}$ with
data and redundant subcarriers, can be constructed as:
\begin{equation} \label{eq:matrix1uw}
\bold{D} =  \bold{P} \left[\begin{array}{c} \bold{d}_d \\ \bold{d}_r \end{array} \right], 
\end{equation}
where $\bold{P} \in \{ 0,1\}^{(N_d+N_r)\times (N_d+N_r)}$ is a permutation matrix, $\bold{d}_d \in \mathbb{C}^{N_d \times M}$ and $\bold{d}_r \in \mathbb{C}^{N_r \times M}$ are the matrices of data and redundant subcarriers, respectively.

The GFDM $I$ and $Q$ 
can be organized in a modulation matrices structured according to:
\begin{equation} 
 \bold{A}^{(.)}=  [\bold{g}_{0,0}^{(.)}~\cdots, \bold{g}_{K-1,0}^{(.)}~\bold{g}_{0,1}^{(.)}~\cdots~\bold{g}_{K-1,M-1}^{(.)}],\\
\end{equation}
where $\bold{g}_{k,m}^{(.)}$ 
is a column vector containing the samples from $g_{k,m}^{(.)} [n],$ 
with $^{(.)}$ equal to $^{(i)}$ or $^{(q)}$ in (\ref{eq:filter}).

The matrices $\bold{A}^{(i)}$ and $\bold{A}^{(q)}$ have dimensions $KM$x$KM$ with an approximately  diagonal block structure, according to:
\begin{equation} \label{eq:matrixGFDM1}
\bold{A}^{(.)}= \left[\begin{array}{lr} [\bold{A}^{(.)}]_1~~ \bold{0} ~~~~~~ \cdots ~~ \bold{0} \\ \bold{0} ~~~~~~ [\bold{A}^{(.)}]_2 ~~ \cdots ~~ \bold{0} \\ \vdots ~~~~~~~ \vdots ~~~~~~~\vdots ~~~~~~ \vdots \\ \bold{0} ~~~~~~ \bold{0} ~~~~~~ \cdots ~~ [\bold{A}^{(.)}]_M \end{array} \right],
\end{equation}
where $[\bold{A}^{(.)}]_m \in \mathbb{C}^{(K)\times (K)}$. 

With these observations, we can write each symbol of the transmitted GFDM block as:
\begin{equation} 
\bold{x}_m = \bold{x}_m^{(i)} + \bold{x}_m^{(q)},
\end{equation}
where
\begin{equation} \label{eq:matrix2uw}
\begin{array}{lr}
\bold{x}_m ^{(i)}=[\bold{A}^{(i)}]_m \bold{P}\Re \left\{ \left[\begin{array}{c} \bold{d}_d(m) \\ \bold{d}_r(m) \end{array} \right]\right\} = \left[\begin{array}{c} \bold{x}_d^{(i)} (m)\\ \bold{0} \end{array} \right],\\
\bold{x}_m ^{(q)}=[\bold{A}^{(q)}]_m \bold{P} \Im \left\{\left[\begin{array}{c} \bold{d}_d(m) \\ \bold{d}_r(m) \end{array} \right]\right\} = \left[\begin{array}{c} \bold{x}_d^{(q)} (m)\\ \bold{0} \end{array} \right].\\
\end{array}
\end{equation}

Writing each result of $[\bold{A}^{(.)}]_m \bold{P}$ with four sub-matrices $\left(\bold{M}_{11}^{(.)}(m),\bold{M}_{12}^{(.)}(m),\bold{M}_{21}^{(.)}(m),\bold{M}_{22}^{(.)}(m)\right)$, it is possible to write $\bold{d}_r^{(.)}(m) = \bold{T}^{(.)}(m)\bold{d}_d^{(.)}(m)$, where $\bold{T}^{(.)}(m) = -\bold{M}_{22}^{(.)}(m) ^{-1} \bold{M}_{21}^{(.)}(m)$.

Due to the real-orthogonality between the matrices $\bold{A}^{(i)}$ and $\bold{A}^{(q)}$, i.e., $\Re \{(\bold{A}^{(i)})^H\bold{A}^{(q)}\}=\Re \{(\bold{A}^{(q)})^H\bold{A}^{(i)}\}=\bold{0}_{MK\times MK}$, a code word $\bold{c}(m)$ can be obtained for each symbol $m$ of the GFDM block by
\begin{equation} \label{eq:matrix3uw}
\begin{array}{l}
\bold{c}(m) =  \bold{P} \left[\begin{array}{c} \bold{d}_d (m)\\ \bold{d}_r (m)\end{array} \right] \approx  \bold{P} \left[\begin{array}{c} \bold{I} \\ \bold{T} (m)\end{array} \right] \bold{d}_d (m) \\~~~~~~= \bold{R} (m)\bold{d}_d(m),
\end{array}
\end{equation}
where $\bold{I}$ is the identity matrix and  $\bold{R} (m)=\bold{P} \left[\begin{array}{c} \bold{I} \\ \bold{T} (m)\end{array} \right]$ can be interpreted as a UW code generator matrix. Another interpretation is that $\bold{R} (m)$ introduces correlations in the vector $\bold{d}(m)$ of frequency domain samples of a GFDM block.

In the second step, $\bold{x}_u(m)$ is added to the GFDM frame. 
The choice of $\bold{x}_u$(m) is made to optimize particular needs in GFDM systems, like synchronization, system parameter estimation purposes or BER gain. The symbols of $\bold{x}_u$ can be placed in one or more symbols of a GFDM block, in a specific pattern, in order to optimize the UW autocorrelation properties. The UW sequences can be obtained by many ways, among which we highlight: the generalized Barker sequence\cite{Golomb} and a CAZAC sequence (Constant Amplitude, Zero Autocorrelation) \cite{Popovic} that are often used for channel estimation, frequency offset estimation and timing synchronization (not investigated in this work); or 
$\bold{x}_u(m)=\bold{0}$, which introduces a systematic complex valued block code structure within the sequence of subcarriers. Note that the gain due to the exploitation of correlations in frequency domain can be regarded as coding gain. 

At the receiver, after performing the equalization and the
time-frequency conversion, observing the influence of the UW on the
received symbol, we can write the received vector as:
\begin{equation} \label{eq:signaluwreceived}
\widetilde{\widehat{\bold{d}}}(m)=\widehat{\bold{c}}(m) + \widetilde{\bold{x}}_u(m) + \widetilde{\bold{\omega}},
\end{equation}
where $\widetilde{\widehat{\bold{d}}}(m) \in \mathbb{C}^{(N_d + N_r)\times 1}$, 
$\widehat{\bold{c}}(m)  \in \mathbb{C}^{(N_d + N_r) \times 1}$ and $\widetilde{\bold{x}}_u(m) \in \mathbb{C}^{(N_d + N_r )\times 1}$.

To eliminate the influence of the UW by subtracting
$\widetilde{\bold{x}}_u(m)$ from
$\widetilde{\widehat{\bold{d}}}(m)$, we can consider that the
channel matrix $\bold{\tilde{H}}$ or at least an estimate of do
$\widehat{\bold{d}}(m)=\widetilde{\widehat{\bold{d}}}(m)-\widetilde{\bold{x}}_u(m)$
such that $\widehat{\bold{d}}(m)=\widehat{\bold{c}}(m) +
\widetilde{\bold{\omega}}$, we can apply the Bayesian Gauss-Markov
theorem~\cite{Kay} to minimize the cost function $J(\epsilon)=
|\epsilon|^2$, with $\epsilon=\bold{c}(m)-\widehat{\bold{c}}(m)$,
obtaining the LMMSE estimator for $\bold{c}(m)$ as given by
 \begin{equation} \label{eq:lmmse}
\widehat{\bold{c}} (m)= \bold{C}_{\bold{c}(m)\bold{c}(m)} \left( \bold{C}_{\bold{c}(m)\bold{c}(m)} + \bold{C}_{\widetilde{\bold{\omega}}\widetilde{\bold{\omega}}}\right)^{-1}\widehat{\bold{d}} (m),
\end{equation}
with the noise covariance matrix $\bold{C}_{\widetilde{\bold{\omega}}\widetilde{\bold{\omega}}} = E\left[\widetilde{\bold{\omega}}\widetilde{\bold{\omega}}^H\right]=K\sigma_{\omega}^{2}\bold{I}$, and with $\bold{C}_{\bold{c}(m)\bold{c}(m)} = E\left[\bold{c}(m)\bold{c}(m)^H\right]=\sigma_{d}^{2}\bold{R} (m)\bold{R} (m)^H$. Recalling that $\bold{c}(m)= \bold{R}(m)\bold{d}_d(m)$, the estimated data vector can be written as 
\begin{equation} \label{eq:wienersmoothing}
\begin{array}{l}
\widehat{\bold{d}}_{d} (m)=\bold{Q}(m)\widehat{\bold{d}} (m),
\end{array}
\end{equation}
where $\bold{Q}(m)$ represents a Wiener smoothing matrix given by
\begin{equation} \label{eq:wiener}
\begin{array}{l}
\bold{Q} (m)= \left(\bold{R} (m)^H\bold{R} (m) + \frac{K\sigma_{n}^2}{\sigma_{d}^2}\bold{I}\right)^{-1} \bold{R}(m)^H .
\end{array}
\end{equation}

The smoothing operation exploits the correlations between subcarrier
symbols which have been introduced by (\ref{eq:matrix3uw}) at the
transmitter and act as a noise reduction operation on the
subcarriers. Other interference cancellation techniques
\cite{delamare_mber,rontogiannis,delamare_itic,stspadf,choi,stbcccm,FL11,delamarespl07,jidf,jio_mimo,peng_twc,spa,spa2,jio_mimo,P.Li,jingjing,memd,did,bfidd,mbdf,bfidd,mserrr,shaowcl08}.
 can also be
examined for receive processing.

The design of the pulse shaping filter of GFDM strongly influences the spectral properties of the signal and the error rate. In this work, we use the Root Raised Cosine (RRC) with the Meyer auxiliary function proposed in~\cite{Nicola} and described by
\begin{equation} \label{eq:rrc}
G \left[f\right]= \sqrt{ \frac{1}{2}\left[1-\cos\left(\pi f\left(\frac{k}{\alpha K}\right)\right)\right]},
\end{equation}
where $f\left(\frac{k}{\alpha K}\right)$ is a truncated function that is used to describe the roll-off area defined by $\alpha$ in the frequency domain. The $k$th subcarrier is centered at the normalized frequency $k/K$ and $\alpha$ describes the overlap of the subcarriers. The function used here is the Meyer auxiliary function $f\left(\frac{k}{\alpha K}\right)=\left(\frac{k}{\alpha K}\right)^4 \left(35-84\left(\frac{k}{\alpha K}\right)+70\left(\frac{k}{\alpha K}\right)^2-20\left(\frac{k}{\alpha K}\right)^3\right)$\cite{Meyer}. 
The use of this function as the argument of a RRC pulse shape
defined in time improves spectrum properties \cite{Gaspar}.


The only increase of the computational complexity at UW-GFDM in
comparison to the CP-GFDM is the computation of the codeword
$\bold{c}(m)$ by (\ref{eq:matrix3uw}), whose computational cost is
$\mathcal{O} (K N_d)$ and in the smoothing $\bold{Q} (m)$ by
(\ref{eq:wienersmoothing}), with $\mathcal{O} (N_d^2)$.

\section{Simulations results}
\label{SimResult}

To validate the UW-GFDM proposal, simulations were performed comparing its performance with the CP-GFDM concept and with the UW-OFDM and CP-OFDM systems, whose performances are already widely known. 
In this work, we consider perfect synchronization, perfect knowledge of the channel by the receiver and $\bold{x}_u(m)=\bold{0}$. 
The simulations were performed with the parameters listed in Table~\ref{parameters}.

\begin{table}[h]
\centering
\caption{Simulation Parameters for the GFDM Systems}
\label{parameters}
\vspace{-.5 em}
\begin{tabular}{|l|c|c|}
  \hline \hline
 Systems & UW-GFDM & CP-GFDM\\ \hline
 Number of subcarriers [K] & 64 & 64\\ \hline
 Data subcarriers [$N_d$]& 48 & 64\\ \hline
 Redundant subcarriers [$N_r$]& 16 & 0\\ \hline
 Length of UW and CP [$L$]& 16 & 16\\ \hline
 Number of time slots [ M] & 4 & 4\\ \hline
 Pulse shaping filter [G] & RRC & RRC\\ \hline
 Roll-off factor [$\alpha$] & $\{0.1, 0.5\}$ & $\{0.1, 0.5\}$\\ \hline
 Modulation subcarriers& FS-OQAM & FS-OQAM\\
  \hline  \hline
  \end{tabular}
  \end{table}

 The CP-OFDM and UW-OFDM are simulated with the same parameters as CP-GFDM and UW-GFDM, respectively, except for the pulse shaping filter that is not used. 
 The indexes of the 16 redundant subcarriers are chosen as in~\cite{Huber}. This choice, with an appropriately constructed matrix $\bold{P}$, provides minimum energy on the redundant subcarriers on average.



\begin{figure}[!h]
\centering
\includegraphics[width=.82\columnwidth]{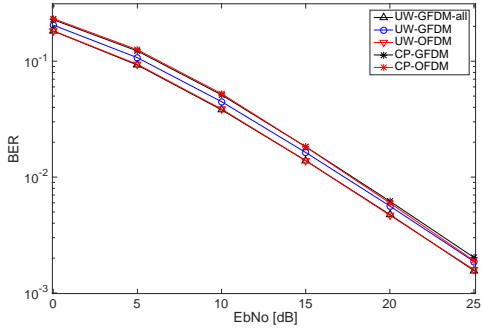}
\vspace{-1em}
\caption{\itshape{BER performance of GFDM and OFDM system with Rayleigh channel}}
\label{ber_performance}
\end{figure}

 In the first example, we observe the performance in a channel that typically represents the WRAN scenarios~\cite{Kim}. These scenarios should be investigated for 5G systems. The channel has been modeled as a tapped delay line, whose power delay profile is ($0, -7, -15, -22, -24, -19$)[dB] for the delays ($0, 3, 8, 11, 13, 21$)[$\mu$s], each tap with Gaussian distribution, and is considered constant during a GFDM block period. In order to provide a simple way of equalization, the length of UW and the CP was set to $L=16$ samples. In Fig.~\ref{ber_performance}, we can observe that the curves UW-GFDM-all and UW-OFDM show a similar performance and approximately 1.2 dB in $E_b/N_0$ better than CP-OFDM and CP-OFDM. It is also possible to observe that the use of a single UW in the first symbol of the GFDM block already reduces the required $E_b/N_0$ by approximately 0.5 dB for the same BER. 


\begin{figure}[!h]
\centering
\includegraphics[width=.82\columnwidth]{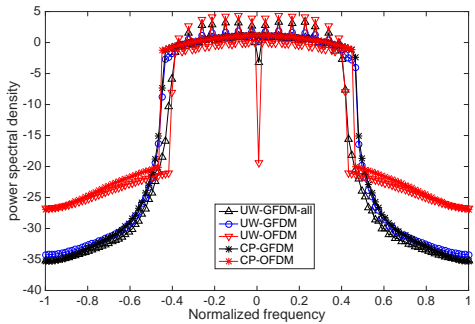}
\vspace{-1em}
\caption{\itshape{OOB performance of GFDM and OFDM system}}
\label{oob_performance}
\end{figure}

In the second example, we can observe that the out-of-band (OOB) emission is not degraded by using the UW, and GFDM maintains its performance around 10 dB better than OFDM, as it had been observed in CP-GFDM and shown in Fig.~\ref{oob_performance}.

An important observation is that the peak-to-average power ratio (PAPR) performance of UW-GFDM is comparable to CP-OFDM and CP-GFDM systems using FS-OQAM modulation.

\begin{figure}[!h]
\centering
\includegraphics[width=.82\columnwidth]{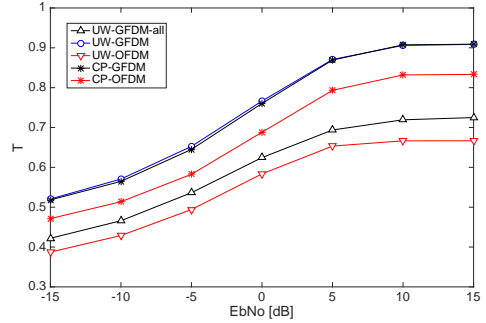}
\vspace{-1em}
\caption{\itshape{Throughput performance of GFDM and OFDM system}}
\label{throughput}
\end{figure}

In the last example, we compare the normalized throughput ($T$) of the systems, calculated by
\begin{equation}
T= \frac{N_d}{K+L}(1-BLER),
\end{equation}
where $BLER$ is the block error rate. 
These comparisons allow us to observe how much the inclusion of redundant subcarriers affects the throughput in relation to the system with CP. 
Fig.~\ref{throughput} shows that the CP-GFDM and UW-GFDM have a comparable throughput (UW-GFDM is slightly better for low $E_b/N_0$ values), and both are $8\%$ higher than CP-OFDM. Observing the curves UW-GFDM-all and UW-OFDM, we can conclude that the use of UW in all sub-symbols of the GFDM block reduces the throughput significantly.

\section{Conclusions}
\label{Conclus}
In this work, we have introduced the concept of UW in GFDM systems. The guard interval is built by a UW instead of cyclic prefixes. 
This approach significantly  reduces the noise on the subcarriers, maintaining the OOB obtained with the pulse shape filter in GFDM, and increases the throughput  as compared to UW-OFDM and CP-OFDM, if applied to only one or a few block sub-symbols. The use of UW in GFDM requires the application of a modulation that allows orthogonality between the subcarriers, and the possibility of the UW for synchronization and channel estimation purposes requires the use of a nonzero UW that is orthogonal to the subcarriers of the GFDM sub-symbol. The proposed UW-GFDM approach outperforms CP-GFDM, UW-OFDM and CP-OFDM in all examples.

\bibliographystyle{IEEEtran}
\bibliography{refsrrpca}

\end{document}